\documentclass[twocolumn,pra,groupedaddress,nofootinbib]{revtex4-1}%
\pdfoutput=1
\usepackage{amsfonts}
\usepackage{color}
\usepackage{amsmath}
\usepackage{amssymb}
\usepackage{graphicx}%
\usepackage[hyperfootnotes=false]{hyperref}
\definecolor{darkblue}{rgb}{0,0,0.5}
\hypersetup{
colorlinks=true,
linkcolor=black,
filecolor=blue,
citecolor=darkblue,  
urlcolor=black,
}
\providecommand{\U}[1]{\protect\rule{.1in}{.1in}}

\DeclareMathOperator*{\argmax}{arg\,max}
\newcommand{\dif}{\mathop{}\!\mathrm{d}}

\newcommand{\nt}{N_B} 
\renewcommand{\d}{\mathrm{d}}
\newcommand{\atermM}{P(k|0;M)}
\newcommand{\btermM}{P(k|\pi;M)}
\newcommand{\ns}{N_S}

\begin{document}

\title{Infinite-fold enhancement in communications capacity using pre-shared entanglement}

\author{Saikat Guha}
\author{Quntao Zhuang}
\author{Boulat Bash}
\affiliation{College of Optical Sciences, University of Arizona, Tucson AZ 85721}
\affiliation{Department of Electrical Engineering, University of Arizona, Tucson AZ 85721}

\begin{abstract}
Pre-shared entanglement can significantly boost communication rates in the regime of high thermal noise, and a low-brightness transmitter. In this regime, the ratio between the entanglement-assisted capacity and the Holevo capacity, the maximum reliable-communication rate permitted by quantum mechanics without any pre-shared entanglement as a resource, is known to scale as $\log(1/N_S)$, where $N_S \ll 1$ is the mean transmitted photon number per mode. This is especially promising in enabling a large boost to radio-frequency communications in the weak-transmit-power regime, by exploiting pre-shared optical-frequency entanglement, e.g., distributed by the quantum internet. In this paper, we propose a structured design of a quantum transmitter and receiver that leverages continuous-variable pre-shared entanglement from a downconversion source, which can harness this purported infinite-fold capacity enhancement---a problem open for over a decade. Finally, the implication of this result to the breaking of the well-known {\em square-root law} for covert communications, with pre-shared entanglement assistance, is discussed.

\end{abstract}
\maketitle

{\em Introduction}---There is much interest in recent years in architecting the {\em quantum internet}~\cite{Kim08, Weh18}, a global network built using quantum repeaters~\cite{Guh15, Mur16} that can distribute entanglement at high rates among multiple distant users per application demands~\cite{Das18,Pan19,Pir19}. There are several well-known applications of shared {\em entanglement}, a new information currency: distributed quantum computing~\cite{Met16}, secure communications with physics-based security~\cite{E91}, provably-secure access to quantum computers on the cloud~\cite{Chi05}, and entanglement-enhanced distributed sensors~\cite{Zhu17a,Pro18,Guo19,Xia19}. In this paper, we elucidate a system design for a yet-another high-impact application of shared entanglement: that of providing a large boost to classical (e.g., radio-frequency, or RF) communication rates.

Transmission of electromagnetic (EM) waves in linear media, as in optical fiber, over the atmosphere or in vacuum, can be described as propagation of a set of mutually-orthogonal spatio-temporal-polarization modes over the single-mode lossy Bosonic channel ${\cal N}_\eta^{N_B}$, described by the Heisenberg evolution ${\hat a}_{\rm out} = \sqrt{\eta}\, {\hat a}_{\rm in} + \sqrt{1-\eta}\,{\hat a}_{\rm E}$, where $\eta \in (0, 1]$ is the modal (power) transmissivity, and the environment ${\hat a}_{\rm E}$ is excited in a zero-mean thermal state of mean photon number per mode $N_B$. Alice encodes classical information by modulating the state of the ${\hat a}_{\rm in}$ modes, with the constraint of $N_S$ mean photons transmitted per mode. The quantum limit of the classical communication capacity, known as the {\em Holevo capacity}, in units of bits per mode, is given by:
\begin{equation}
C(\eta, N_S, N_B) = g(N_S^\prime) - g((1-\eta)N_B),
\end{equation}
where $N_S^\prime \equiv \eta N_S + (1-\eta)N_B$ is the mean photon number per the ${\hat a}_{\rm out}$ mode at the channel's output received by Bob, and $g(x) \equiv (1+x)\log(1+x) - x\log(x)$ is the von Neumann entropy of a zero-mean single-mode thermal state of mean photon number $x$~\cite{Gio04,Gio14}~\footnote{All logarithms in this paper are taken to base $2$. The symbol $\ln$ is used for natural logarithm.}.

If Alice and Bob pre-share (unlimited amount of) entanglement as an additional resource, but operating under the same conditions as above---transmitting classical data over ${\cal N}_\eta^{N_B}$ with a transmit photon number constraint of $N_S$ photons per mode---the capacity, in units of bits per mode, increases to the following~\cite{BSST,Gio03,Hol01,Hol02,Hol03,Hol04}:
\begin{equation}
C_E(\eta, N_S, N_B) = g(N_S) + g(N_S^\prime) - g(A_+) - g(A_-),
\end{equation}
where $C_E$ is the {\em entanglement assisted classical capacity} of the quantum channel ${\cal N}_\eta^{N_B}$, and
$
A_{\pm} = \frac12(D-1 \pm (N_S^\prime - N_S)), 
$
with
$
D = \sqrt{(N_S + N_S^\prime + 1)^2 - 4\eta N_S(N_S+1)}.
$

In the regime of a low-brightness transmitter ($N_S \ll 1$) and high thermal noise ($N_B \gg 1$),
\begin{equation}
\frac{C_E}{C} \approx \ln\left(\frac{1}{N_S}\right),
\label{eq:scaling}
\end{equation}
which goes to infinity as $N_S \to 0$~\cite{Shi19}. The practical implication of this can be potentially revolutionary in RF communications, since the condition $N_B \gg 1$ is naturally satisfied at the longer center wavelengths characteristic of RF. Exploiting (optical frequency) pre-shared entanglement between Alice and Bob---distributed via a repeatered quantum internet---potentially an order of magnitude or more enhancement in classical communications rate is possible, depending upon the actual operational regime of loss, noise, and transmit power, compared to conventional RF communications that does not use pre-shared entanglement as a resource. See Supplementary Information for a more quantitative discussion on this.

Despite the large capacity advantage attainable with pre-shared entanglement been known for decades, a structured transmitter-receiver design to harness this enhancement has eluded us. Continuous-variable (CV) superdense coding yields a factor-of-two capacity advantage in the noiseless case, but does not provide any advantage in the noisy regime~\cite{Soh03}. It was recently shown that phase-only encoding on pre-shared two-mode squeezed vacuum states attains $C_E$ in the $N_S \ll 1$, $N_B \gg 1$ regime~\cite{Shi19}, but with a receiver measurement that does not translate readily to a structured optical design. Receivers based on optical parametric amplification (OPA)~\cite{Guh09} and sum-frequency-generation (SFG)~\cite{Zhu17} only provide at most a factor-of-$2$ improvement over $C$, as shown in~\cite{Shi19} and the Supplementary Information.

In this paper, we take an important step towards solving this long-standing open problem. We combine insights from the SFG receiver proposed for a quantum illumination radar~\cite{Zhu17}, and the Green Machine (GM) receiver proposed for attaining superadditive communication capacity with phase modulation of coherent states~\cite{Guh11b}~\footnote{Jet Propulsion Laboratory developed a decoding algorithm for the first-order length-$n$ Reed Muller codes that employed the fast Hadamard transform in a specialized circuit that used $(n\log n)/2$ symmetric buttery circuits, for sending images from Mars to the Earth as part of the Mariner 1969 Mission. This circuit came to be known as the {\em Green Machine} named after its JPL inventor. Guha developed an optical version of the Green Machine decoding circuit, replacing the butterfly elements by 50-50 beamsplitters, which he showed achieved superadditive communication capacity with Hadamard-coded coherent-state BPSK modulation, i.e., communication capacity in bits transmissible reliably per BPSK symbol that is fundamentally higher than that is physically permissible with any receiver that detects each BPSK modulated pulse one at a time~\cite{Guh11b}. This paper's joint detection receiver for entanglement assisted communications leverages insights from that optical Green Machine.}, to obtain a transmitter-modulation-code-receiver structured design that saturates the $\ln(1/N_S)$ scaling in capacity gain over the Holevo capacity in~\eqref{eq:scaling}.


\begin{figure}
\centering
\includegraphics[width=\columnwidth]{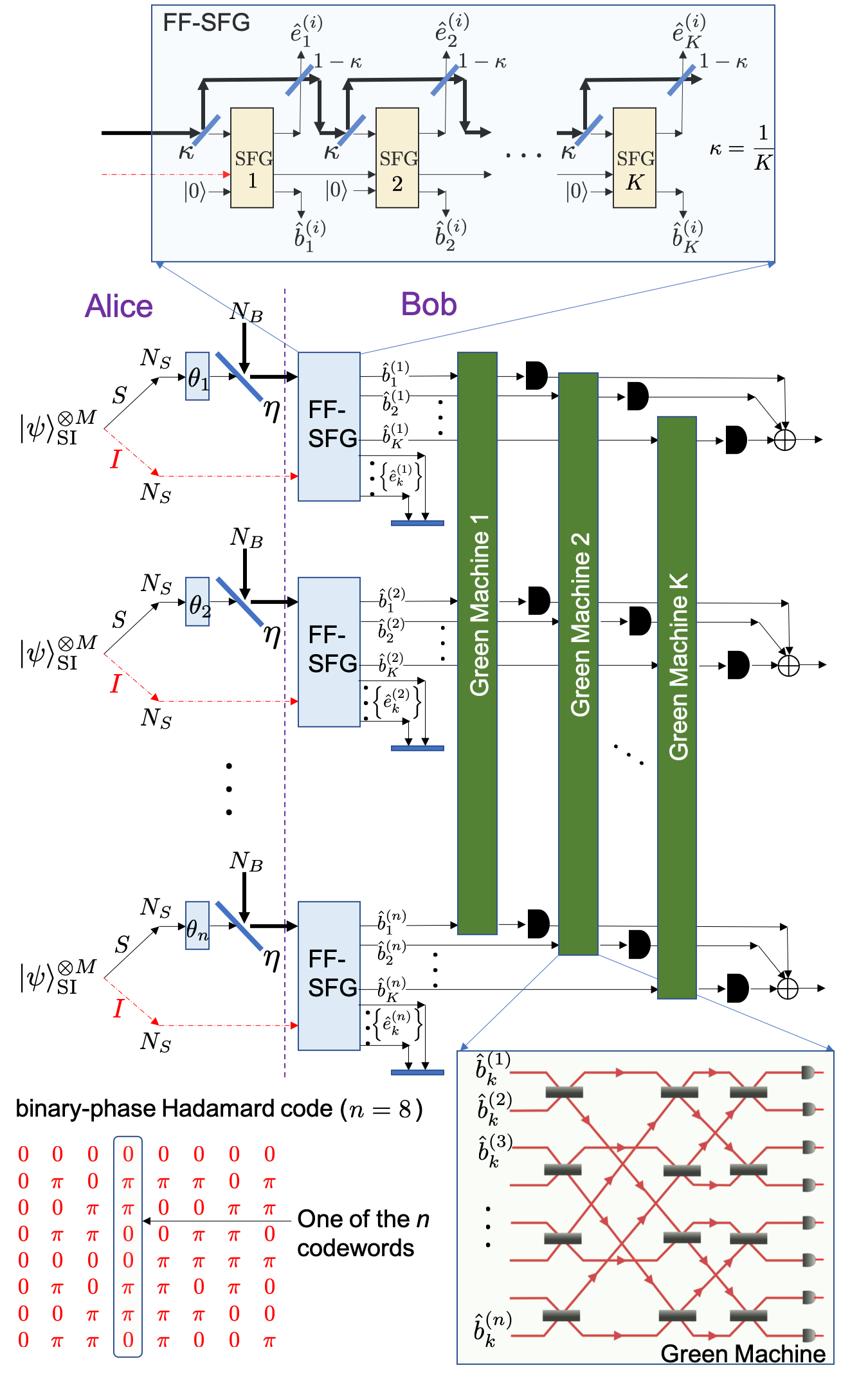}
\caption{A schematic of our joint detection receiver, which acts on $Mn$ signal modes modulated by Alice, which are received by Bob after transmission through the lossy-noise Bosonic channel ${\cal N}_\eta^{N_B}$, and $Mn$ idler modes held by Bob, entangled with Alice's transmitted modes. The pre-shared entanglement is shown using red (dash-dotted) lines. In an actual realization, only one $n$-mode Green Machine is needed, because the sum-frequency modes ${\hat b}_k^{(i)}$, $1 \le k \le K$ come out in a temporal sequence.}
\label{fig:JDR}
\end{figure}
{\em Joint detection receiver design}---Let us consider the transmitter-receiver structure sketched in Fig.~\ref{fig:JDR}. Alice employs a binary phase shift keying (BPSK) modulation with a Hadamard code of order $n$. Let us assume $n$ is a power of $2$ such that a Hadamard code exists. A block of $M$ temporal modes of the signal output of a pulsed spontaneous parametric downconversion (SPDC) source, an $M$-fold tensor product two-mode squeezed vacuum $|\psi\rangle_{\rm SI}^{\otimes M}$, is modulated by one value of binary phase $\theta_i \in \left\{0, \pi\right\}$. The transmission of an entire BPSK-modulated Hadamard codeword consumes $n$ SPDC signal pulses, modulated with phases $\theta_i, 1 \le i \le n$, consuming $nM$ uses of the single-mode channel ${\cal N}_\eta^{N_B}$. The corresponding idler modes are losslessly pre-shared with Bob, e.g., using a fault-tolerant quantum network. Alice's phase modulation of the signal modes, followed by transmission of the signal modes through ${\cal N}_\eta^{N_B}$, turns into phase modulation of (classical) phase-sensitive cross correlations between Bob's received modes and (losslessly-held) idler modes. This correlation bears the information in Alice's phase modulation through the lossy-noisy channel much stronger than any classical means, e.g., an amplitude-phase modulated coherent state.

To translate phase modulation of phase-sensitive signal-idler cross correlations into modulation of (quadrature) field displacement, for which we have significant prior literature on receiver designs, e.g., for phase modulated coherent states, we employ SFG, a non-linear optical process, which runs SPDC in reverse, per the Hamiltonian ${\hat H}_I = \hbar g \sum_{m=1}^M \left({\hat b}^\dagger {\hat a}_{S_m} {\hat a}_{I_m} + {\hat b} {\hat a}_{S_m}^\dagger {\hat a}_{I_m}^\dagger\right)$, with $\hbar$ the reduced Planck constant, and $g$ the non-linear interaction strength. Signal-idler photon pairs from the $M$ input mode pairs are up-converted to a sum-frequency mode $\hat b$, and the phase-sensitive cross-correlations $\langle {\hat a}_{S_m}{\hat a}_{I_m} \rangle$ manifests as a (quadrature) displacement of a thermal state of the $\hat b$ mode~\cite{Zhu17}.

Bob employs $n$ feed-forward (FF) SFG modules---made by stacking $K$ SFG stages, each of duration $\pi/2\sqrt{M}g$, and $K$ beamsplitters and combiners of transmissivities $\kappa = 1/K$ and $1-\kappa$ respectively, as shown in Fig.~\ref{fig:JDR}---to mix the $nM$ modulated-received modes with the $nM$ locally-held idler modes, pre-shared with Bob, entangled with Alice's signal modes. The reason for the $K$-stage SFG is that the bright noise background results in bright received modes, and that we wish the signal input of each SFG stage to have much less than a photon per mode, so that we can borrow the ``qubit-approximation" analysis of the SFG from~\cite{Zhu17}. ${\hat b}_k^{(i)}$ denotes the sum-frequency mode of the $k$-th SFG, $1 \le k \le K$, of the $i$-th FF-SFG module, $1 \le i \le n$. The sum-frequency outputs ${\hat b}_k^{(i)}$, $1 \le i \le n$ from the $K$ FF-SFG modules are input into an $n$-mode linear-optical Green Machine (GM) circuit GM$_k$, each of which has $n$ outputs that are each detected by single photon detectors~\cite{Guh11b}. An $n$-mode GM, as shown in the bottom right of Fig.~\ref{fig:JDR}, is a linear-optical circuit comprising $n\log_2(n)/2$ $50$-$50$ beasmplitters. It turns an $n$-mode BPSK-modulated coherent-state Hadamard codeword at its input into one of the $n$ codewords of an order-$n$ coherent-state pulse-position modulation (PPM) at its output. The electrical outputs of the $i$-th detectors from each of the $K$ GM modules are classically combined into one output that is monitored for zero or more clicks, during each SPDC pulse interval. Since the $K$ sum-frequency modes ${\hat b}_k^{(i)}$, $1 \le k \le K$ in the $i$-th FF-SFG module come out in a temporal sequence, in reality we will only need one $n$-mode GM and $n$ detectors. The diagram in Fig.~\ref{fig:JDR} shows $K$ GMs for ease of explanation.

Define ${\hat \rho}_{\rm th}(\alpha, N_T) = \int_{\mathbb C} \frac{1}{\pi N_T}e^{-(\beta - \alpha)^2/N_T}|\beta\rangle \langle \beta | d^2\beta$ as a single-mode thermal state with mean field amplitude $\alpha \in {\mathbb C}$. The photodetection statistics of this state is Laguerre-distributed~\cite{Hel76}. The probability that this produces zero clicks when detected with an ideal photon detector, $\langle 0|{\hat \rho}_{\rm th}(\alpha, N_T) |0\rangle = (1/(N_T+1))e^{-|\alpha|^2/(N_T+1)}$. In the $\kappa \ll 1/N_B$ limit, for the $k$-th GM, the $n$ input modes are in states ${\hat \rho}_{\rm th}(\pm \alpha^{(k)}, N_T)$, where the $\pm$ signs are governed by the specific Hadamard codeword that was used,  $\alpha^{(k)} = \sqrt{M\kappa \eta N_S(1+N_S)\mu^{k-1}}$, with $\mu = \left(1-\kappa(1+N_S^\prime)\right)^2$, and $N_T = \kappa N_SN_S^\prime$~\cite{Zhu17}. Let us also define $N_k = |\alpha^{(k)}|^2$. One of the $n$ output modes of the $k$-th GM (which one, based on which Hadamard code was sent) is in a displaced thermal state ${\hat \rho}_{\rm th}(\sqrt{n}\,\alpha^{(k)}, N_T)$. We call this the ``pulse-containing output" (mode). The remaining $n-1$ output modes are in the zero-mean thermal state ${\hat \rho}_{\rm th}(0, N_T)$. At the $n$ classically-combined detector outputs---produced by detecting one Hadamard codeword, i.e., $Mn$ received-idler mode pairs---we record a random binary $n$-vector of (no-click, click), i.e., $2^n$ possible outcomes. The $2^n$ click patterns are clubbed into $n+1$ outcomes: a click in a given output and no clicks elsewhere, or an {\em erasure}, which refers to either zero clicks on all $n$ outputs, or multiple clicks in any of the outputs.

The modulation-code-receiver sequence described above induces an $n$-input $n+1$-output discrete memoryless channel (DMC), which happens to be identical to the DMC induced by coherent-state pulse-position modulation (PPM) and single photon detectors with non-zero background (or, dark) click probability. The capacity of this channel~\cite{Jar17}, divided by $(Mn)$, is the bits per mode capacity attained by our modulation-code-receiver trio:
\begin{eqnarray}
R_E^{(M,n)} &=& \frac{1}{Mn}\left( {p_e}\log n + (n-1)p_d\log \frac{np_d}{p_e} \right. \nonumber \\
&-& \left.\big(p_e+(n-1)p_d\big)\log\left[1+\frac{(n-1)p_d}{p_e}\right]\right),
\end{eqnarray}
where $p_d = (1-p_c)p_b(1-p_b)^{n-2}$ and $p_e = p_c(1-p_b)^{n-1}$.

In the above formula, $1 - p_c$ is the probability that the pulse-containing output of the receiver does not produce any clicks, and $1 - p_b$ is the probability that any given non-pulse-containing output does not produce any clicks. Assuming the photodetection statistics of the $i$-th outputs of each of the $K$ GMs are statistically independent, we get $1 - p_c = \Pi_{k=1}^K (1-p_c^{(k)})$, where $1-p_c^{(k)} = \frac{1}{N_T+1}e^{-nN_k/(N_T+1)}$. This simplifies to:
\begin{eqnarray}
p_c &=& 1-\frac{1}{(1+N_T)^K}e^{-A\left(\frac{1-\mu^K}{1-\mu}\right)}, {\text{and}}\\
p_b &=& 1 -\frac{1}{(1+N_T)^K},
\end{eqnarray}
with $A = nM\kappa\eta N_S(N_S+1)/(N_T+1)$.

\begin{figure}
\centering
\includegraphics[width=\columnwidth]{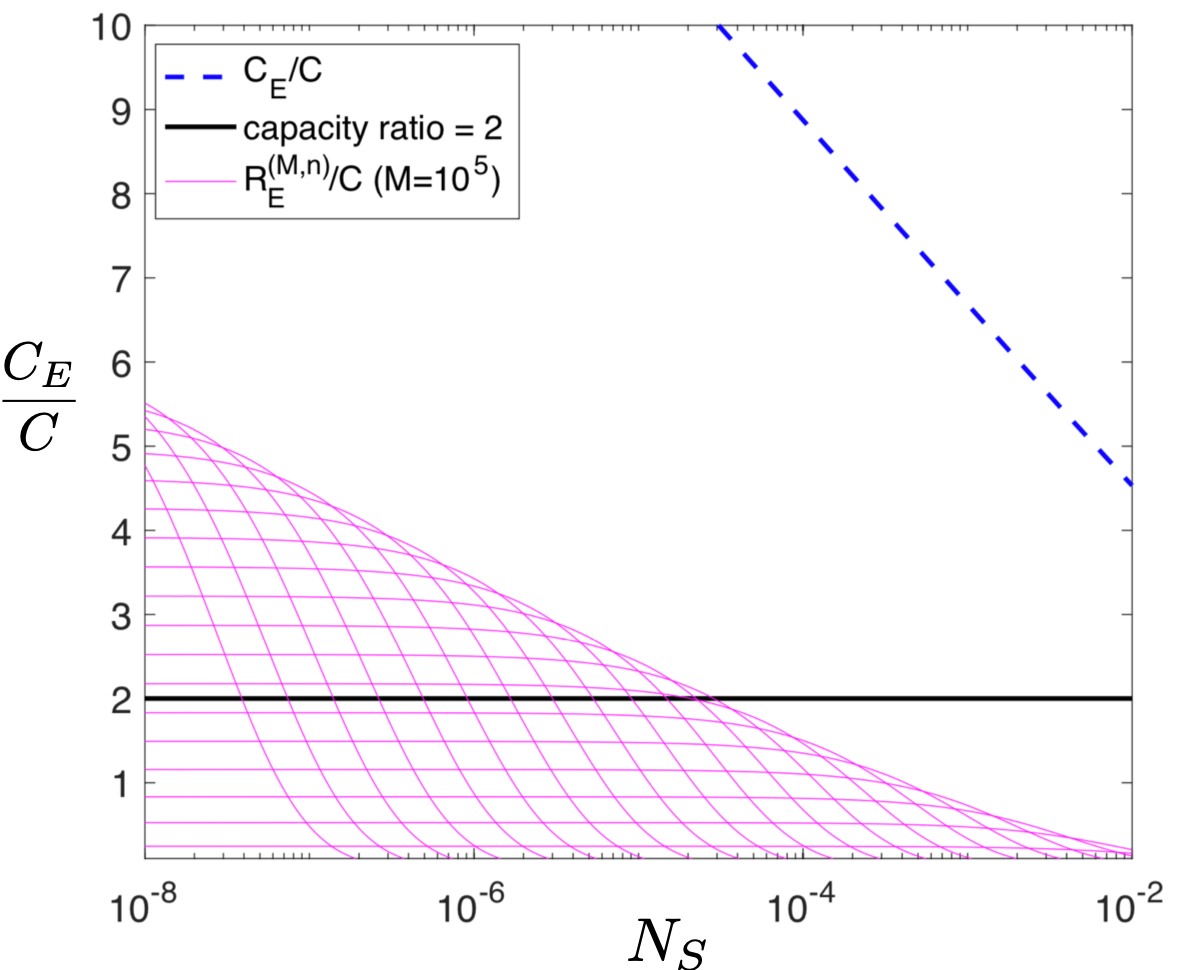}
\caption{The thin magenta lines are plots of $R_E^{(M,n)}/C$ for $n = 2, 4, 8, \ldots, 2^{20}$, for $M = 10^5$. This shows that the capacity ratio scales as $\log(1/N_S)$, which goes to infinity as $N_S \to 0$, for any given $M$. However, this scheme (BPSK modulation, Hadamard code, and our proposed structured joint-detection receiver) does not achieve $C_E$. We have assumed $\eta = 0.01$ and $N_B = 10$ photons per mode, for all the plots in this Figure.}
\label{fig:cap_ratio_M100000}
\end{figure}
In Fig.~\ref{fig:cap_ratio_M100000}, we plot the ratio $C_E/C$ as a function of $N_S$ in the $N_S \ll 1$ regime, for $\eta = 0.01$ and $N_B = 10$. We plot the capacity ratios $R_E^{(M,n)}/C$, attained for $M=10^5$, and $n \in \left\{2, 2^2, \ldots, 2^{20}\right\}$. Let us define $R^{(M)}_E = {{\rm sup}_{n}R_E^{(M,n)}}$ to be the envelope of capacities attained by our scheme over all $n$, for a given $M$.

In order to derive the asymptotic capacity scaling, we apply the conditions pertinent to our problem setting,
\begin{equation}
\eta N_S \ll N_S \ll 1 \ll N_B \ll 1/\kappa,
\end{equation}
and through a series of approximations, and leveraging analytical connections to noisy pulse-position modulation (PPM), we prove in the Supplementary Information:
\begin{equation}
\frac{R_E^{(M)}}{C} \sim \log\left(\frac{1}{N_S}\right),
\label{eq:REM_scaling_optimal}
\end{equation}
establishing that our modulation-code-receiver combination attains the optimal scaling of entanglement-assisted communications in the aforesaid regime, and despite not meeting $C_E$, is in principle capable of harnessing the infinite-fold capacity enhancement possible using shared entanglement---using quantum optical states, processes and detection schemes that are readily realizable. Further, this capacity ratio is clearly larger than $2$, the best achievable ratio with an OPA receiver~\cite{Guh09} or an FF-SFG receiver~\cite{Zhu17,Shi19} (See Supplementary Information). 

{\em Covert communications}---An operational regime that justifies the $N_B \gg 1$ assumption, required for the $\log(1/N_S)$ entanglement-assisted capacity-ratio gain, is radio-frequency (RF), or microwave domain, signaling. Furthermore, aside from practical constraints of the peak source power and high losses, e.g., which may occur in deep turbulent atmospheric propagation or long-range deep-space communications, one obvious regime where $N_S \ll 1$ would be applicable is {\em covert} or provably undetectable communications. Pre-shared entanglement, e.g., distributed at optical frequencies by a future satellite network or the quantum internet, could be leveraged to enhance---potentially by an order of magnitude or more---the amount of information that an RF communication link could transmit provably covertly, i.e., ensuring that the transmission attempt is undiscoverable even by an all-powerful quantum-equipped adversary. For provably covert communications, regardless of whether Alice and Bob employ entanglement assistance or not, the mean transmitted photon number per mode $N_S$ must satisfy $N_S \le \left(\sqrt{2\eta N_B(1+\eta N_B)}/(1-\eta)\right)\, \sqrt{\delta / m}$, where $m$ is the total number of transmitted modes, and $\delta$ quantifies how stringent Alice and Bob are on being covert. The above condition on $N_S$ comes from Alice and Bob setting a requirement that the adversary's probability of error $P_e$, in detecting their transmission attempt must satisfy, $1/2 \ge P_e \ge 1/2 - \delta$. This dependence of $N_S$ on $m$ ultimately leads to the {\em square-root law} of covert communications, i.e., $O(\sqrt{m})$, but no more, bits can be transmitted reliably yet covertly~\cite{Bas15,Bul19}.

Both the OPA and the FF-SFG receivers achieve up to a factor of $2$ enhancement over $C$ (see Supplementary Information, and~\cite{Shi19}). Hence, covert communications using either of those receivers will obey the square-root law, albeit with a factor of $2$ enhancement in the scaling constant. Our scheme in Fig.~\ref{fig:JDR} can achieve a factor of $\log(1/N_S)$ capacity enhancement, in the $N_S \ll 1$, $N_B \gg 1$ regime. This will translate to being able to transmit $O(\sqrt{m}\, \log m)$ bits of information reliably and covertly, thereby breaking the square-root law of covert communications (by leveraging pre-shared entanglement). However, a more careful analysis of this is in order: both to find the constant in the aforesaid scaling, and more importantly to prove a rigorous converse result to provably-covert entanglement-assisted communications. We leave such an analysis of our joint-detection receiver in the covert communication regime, for future work.

{\em Practical considerations and discussion}---For the assumed values of $\eta = 0.01$ and $N_B = 10$ photons per mode, the highest capacity achieved by the joint-receiver receiver discussed above, occurs at around $M \sim 10^5$. A more detailed discussion of why there is an optimal modulation-block length $M$ is discussed in the Supplementary Information. For a typical SPDC entanglement source of optical bandwidth $W \sim$ THz, With $M \sim WT$, $M = 10^5$ translates to a pulse duration $T \sim 100$ ns. This means the BPSK phase-modulation bandwidth necessary would be $\sim 10$ MHz, which is readily realizable with commercial-grade electro-optical modulators at $1550$ nm.

In order to bridge the remainder of the gap to $C_E$, better codes and more complex quantum joint detection receivers will be needed, based on arguments closely aligned with those in~\cite{Chu17}. We believe that the capacity achieved by the receiver in Fig.~\ref{fig:JDR} can be improved by adopting an FF scheme to make use of the extra modes ${\hat e}_k^{m}$'s in Fig.~\ref{fig:JDR}, which was crucial for the optimality of the FF-SFG receiver for quantum radar~\cite{Zhu17}. Further improvement is possible via leveraging insights from a quantum joint-detection receiver for classical optical communications~\cite{Guh11a} which combines the GM and the Dolinar receiver~\cite{Dol73}. This improved scheme would modulate $M$-mode SPDC pulses using a BPSK first-order Reed-Muller code, but now FF-SFG modules will be sandwiched by non-zero-squeezing two-mode-squeezing stages as in~\cite{Zhu17}, and the detectors at the output of the GM stages will feed back into setting the aforesaid squeezing amplitudes, adaptively. We leave this calculation for future work.

It should be obvious that we could have instead used a PPM modulation format, instead of BPSK Hadamard codewords followed by the GM stages, and achieved the same capacity performance. In such a scheme, Alice and Bob would need to pre-share (brighter) SPDC signal-idler mode pairs of mean photon number per mode $nN_S$, and Alice would send an $M$-temporal mode signal pulse (of mean photon number $nN_S$) and nothing (vacuum) in $n-1$ pulse slots. So, only $M$ modes will be excited out of each $Mn$ transmitted modes. FF-SFG stages will be used to demodulate, as before, but no GM stages will be needed. Since the optimal PPM order is $n \sim ({\cal E}\log(1/{\cal E}))^{-1}$ with ${\cal E} = M\eta N_S/(2N_B)$ (see Supplementary Information), which translates to $nN_S \sim \frac{N_0}{\log(N_0/N_S)}$ with $N_0 = 2N_B/(M\eta)$. For the numbers in Fig.~\ref{fig:cap_ratio_M100000}, i.e., $\eta=0.01$, $N_B=10$, $M=10^5$, we get $N_0 = 0.2$. This implies that for $N_S < 0.01$, we get $nN_S \lesssim 0.07$. Thus the idler pulses are still in the regime that the implicit ``qubit approximation" analysis of the SFG borrowed from~\cite{Zhu17} is valid. We relegate a slightly more detailed discussion of PPM and on-off-keying (OOK) modulation for entanglement-assisted communications, to the Supplementary Information. There, we also discuss pros and cons of the BPSK modulation described in this paper, and PPM or OOK modulation, both with regards to the requirements on shared entanglement, and the complexity of the receiver.

It should be further noted that the PPM modulation format in the context of entanglement-assisted communications as described above, was proposed for entanglement-assisted communication over a general quantum channel over a finite-dimensional Hilbert space~\cite{Din17}. This technique has been termed ``position based encoding" in the quantum information theory literature~\cite{Qi18}. However, there is no simple translation known as yet of the receiver measurement that must be employed to achieve $C_E$ with position-based encoding, into a structured optical receiver. It will be interesting, in future work, to find a structured optical receiver design that achieves the full entanglement-assisted capacity $C_E$ afforded by quantum mechanics.

A final point worth noting: pre-shared entanglement affords a large capacity enhancement in the regime of low transmitted signal power per mode and high thermal-noise mean photon number per mode, despite that entanglement does not survive propagation through this (entanglement-breaking) channel. It is this exact same regime where an entangled-state transmitter was shown to attain a superior performance compared to any classical source, for detecting a target at stand-off range---a concept termed {\em quantum illumination}~\cite{Tan08,Guh09,Zhu17}. These two observations are intimately related. These are both tasks that involve extracting information modulated into one half of a two-mode-entangled state where the information-bearing half undergoes propagation over an entanglement-breaking channel. 

{\em Acknowledgments}---SG acknowledges General Dynamics Mission Systems for supporting this research, and works performed under the DARPA Information in a Photon program (2010-2013) under contract number HR0011-10-C-0159, for extremely valuable insights. QZ and BB were sponsored by the Army Research Office under Grant Numbers W911NF-19-1-0418 and W911NF-19-1-0412, respectively. SG acknowledges Ali Cox, Michael Bullock, Christos Gagatsos and Zheshen Zhang for valuable discussions.

\newpage

\appendix

\vspace{10pt}
\begin{center}
{\large{\textbf{Supplementary Information}}}
\end{center}

\section{Bit rate scaling in the low photon number regime}

The purpose of this appendix is to show that in the relevant regime of operation of the entanglement-assisted communication system described in the main paper, i.e.,
\begin{equation}
\eta N_S \ll N_S \ll 1 \ll N_B \ll K,
\label{eq:operationalregime}
\end{equation}
the scaling of the ratio between the rate achieved by our proposed joint detection receiver and the Holevo capacity $R_E^{(M)}/C$ matches the ratio between the entanglement-assisted capacity and the Holevo capacity $C_E/C$:
\begin{equation}
\frac{R_E^{(M)}}{C} \sim \frac{C_E}{C} \sim \ln\left(\frac{1}{N_S}\right),
\label{eq:REM_scaling}
\end{equation}
where $N_S$ is the mean transmitted photon number per mode, and $M$ is the length (number of modes) of the modulated SPDC signal pulse. We first derive $C_E/C$. 

\subsection{Entanglement assisted capacity enhancement}

Intuitively, the scaling $\frac{C_E}{C} \sim \log\left(\frac{1}{N_S}\right)$ in~\eqref{eq:operationalregime} follows from the dominant term in the expression for $C_E$ as $N_S\to 0$ being $-N_S\log N_S$ for any constant $N_B>0$, while the Taylor series expansion of $C$ at $N_S=0$ yielding $C=N_S\log\left(1+((1-\eta)N_B)^{-1}\right)+o(N_S)$.
Formally, one can use L'H\^{o}pital's rule to obtain the following limit:
\begin{align}
\label{eq:CEbyClogNs_limit}\lim_{N_S\to0}\frac{C_E}{C\ln \left(\frac{1}{N_S}\right)}&=\frac{1}{(1+(1-\eta)N_B)\ln\left(1+\frac{1}{(1-\eta)N_B}\right)},
\end{align}
which yields the scaling.  Note that the right hand side (RHS) of \eqref{eq:CEbyClogNs_limit} is zero when $N_B=0$, corresponding to the fact that the ratio $C_E/C\leq2$ in the noiseless regime.

\begin{figure}
\centering
\includegraphics[width=\columnwidth]{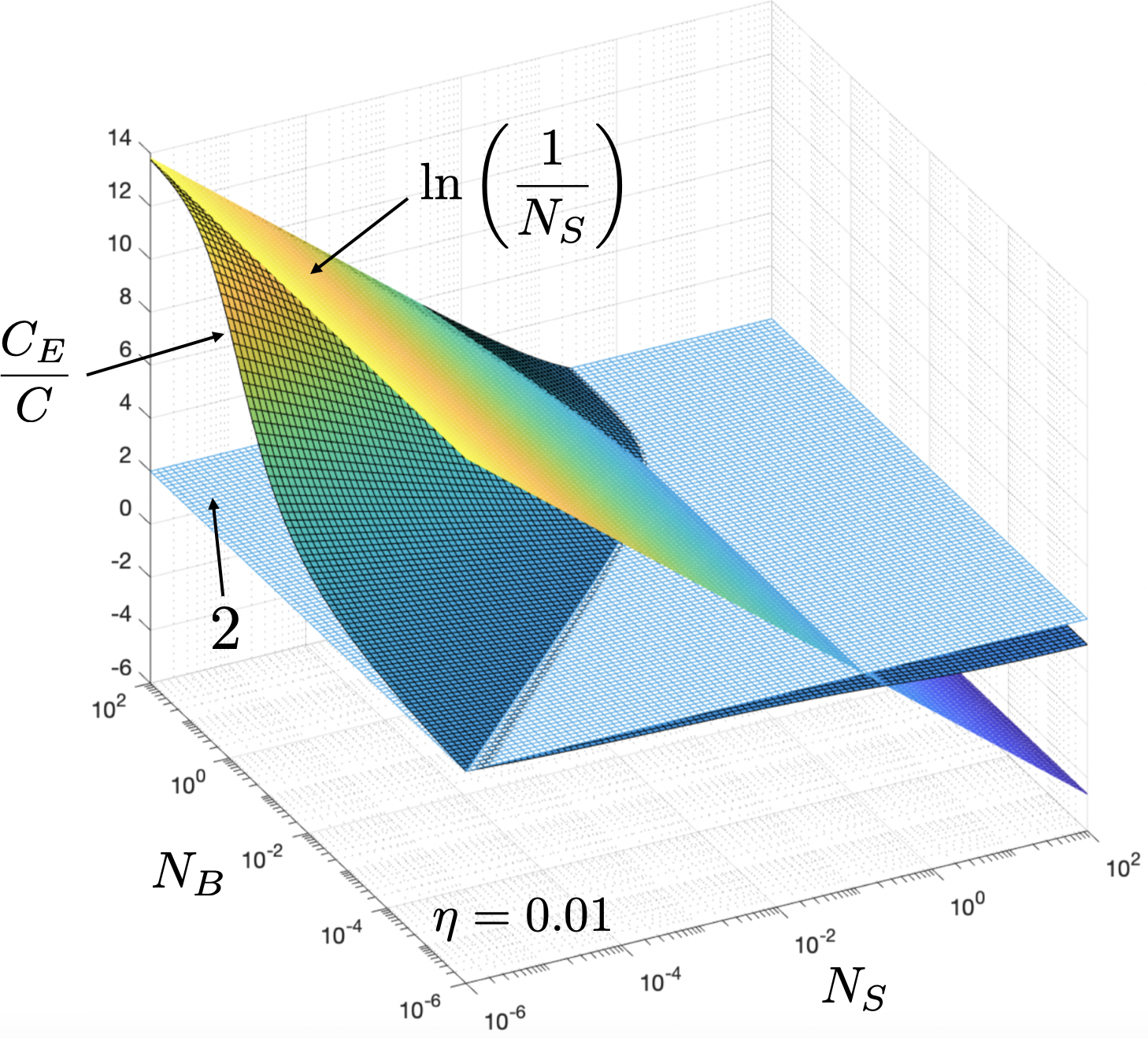}
\caption{The ratio $C_E/C$ as a function of $N_S$ and $N_B$ for channel transmissivity $\eta = 0.01$. Also shown is $\ln(1/N_S)$, which is the scaling of $C_E/C$ as $N_S \to 0$ and $N_B \to \infty$.}
\label{fig:capacity_ratio}
\end{figure}

The plot of the ratio $C_E/C$ as a function of $N_S$ and $N_B$, for channel transmissivity $\eta = 0.01$ in Fig.~\ref{fig:capacity_ratio} yields further insight. At optical frequencies, the Planck-Law limited thermal-noise mean photon number per mode $N_B$ ranges between $10^{-5}$ to $10^{-6}$. At such small $N_B$ values, despite the scaling in \eqref{eq:REM_scaling}, the actual capacity ratio is essentially at or below $2$ (the maximum value when $N_B=0$) over the entire range of chosen values of $N_S$, $10^{-6}$ to $10^2$. The ratio would be significantly large only for the extremely small values of $N_S$ that are not physically meaningful. However, at $N_B = 100$, which is quite reasonable at microwave wavelengths, $C_E/C$ exceeds $10$.

\subsection{Proof of optimal capacity scaling achieved by the joint detection receiver}

Consider the use of an order $n$ pulse position modulation (PPM) scheme over a channel with loss and noise. PPM encodes information by the position of a pulse (e.g., a coherent state of light) in one of $n$ orthogonal modes (e.g., time bins) at the input, which is direct-detected at the output (e.g., by a single photon detector). Loss attenuates the transmitted pulse amplitude, and noise results in potential detection events in one or more bins. Ignoring detection events in multiple bins (i.e., treating them as ``erasures"), and assuming an equiprobable selection over the $n$ inputs (which maximizes the throughput), the Shannon mutual information---expressed in bits per mode---of the $n$-input $n+1$-output discrete memoryless channel (DMC) induced, is given by~\cite[Eq.~(16)]{Jar17}:
\begin{align}
I^{(n)}_{\text{PPM}} &= \frac{p_e}{n}\log n + \frac{(n-1)}{n}p_d\log \frac{np_d}{p_e} \nonumber \\
\label{eq:In}&\phantom{=}- \left[\frac{p_e+(n-1)p_d}{n}\right]\log\left[1+\frac{(n-1)p_d}{p_e}\right],
\end{align}
where $p_e$ is the probability of the detection event occurring exclusively in the bin corresponding to the position of the pulse at the input, and $p_d$ is the probability that a detection event occurs in a single bin that is different from the one containing the input pulse.  Denoting by $p_c$ the probability of a detection event in the bin corresponding to the input pulse and by $p_b$ the probability of a detection event in another bin \cite[Sec.~IV]{Jar17},
\begin{align}
\label{eq:pe}p_e& = p_c(1-p_b)^{n-1}, \,{\text{and}}\\
\label{eq:pd}p_d& = (1-p_c)p_b(1-p_b)^{n-2}.
\end{align}
We specialize the result from \cite{Jar17} to find the channel capacity of the DMC induced by the modulation-code-channel-receiver combination described in Fig.~1 of the main paper. Let us recall that the scheme involves BPSK-modulation of the signal modes of $M$ pre-shared two-mode-squeezed-vacuum (TMSV) states, repeating the above $n$ times, encoding an order-$n$ binary Hadamard code, and transmission of the $Mn$ modulated modes over $Mn$ uses of the single-mode lossy-noisy bosonic channel $\mathcal{N}_{\eta}^{N_B}$, followed by demodulation and detection by our joint detection receiver (JDR). This scheme results in detection events that are statistically identical to demodulating PPM in the presence of noise. Thus, we seek:
\begin{align}
\label{eq:REM}R_E^{(M)}&=\max_{n}\frac{1}{M}I^{(n)}_{\rm PPM},
\end{align}
where we determine $p_e$ and $p_d$ as follows. First, let's recall the definitions. The mean number of photons per mode in the signal modes of the TMSV transmitted by Alice is $N_S$, and the mean photon number of the thermal noise background per transmitted mode is $N_B$. The modal power transmissivity of the bosonic channel is $\eta\in(0,1]$, which implies that Bob's received mean number of photons per mode is $N_S^\prime=\eta N_S+(1-\eta)N_B$. To calculate $p_c$ and $p_b$, we assume the photodetection statistics of the $i$-th outputs of each of the $K$ Green Machines in the JDR are statistically independent, and $K \gg N_B$. Thus, $p_c = 1-\prod_{k=1}^K (1-p_c^{(k)})$, where $1-p_c^{(k)} = \frac{1}{N_T+1}e^{-nN_k/(N_T+1)}$ with $N_T=N_S N_S^\prime/K$, $N_k=\frac{M\eta N_S(1+N_S)\mu^{k-1}}{K}$, and $\mu=\left[1-\frac{1+N_S^\prime}{K}\right]^2$. Thus, we have:
\begin{align}
p_c &= 1-\frac{1}{(1+N_T)^K}e^{-A\left(\frac{1-\mu^K}{1-\mu}\right)}, {\text{and}}\\
p_b &= 1 -\frac{1}{(1+N_T)^K},
\end{align}
with $A = \frac{nM\eta N_S(N_S+1)}{K(N_T+1)}$. Using the conditions:
\begin{align}
N_S \ll 1 \ll N_B \ll K,
\end{align}
we can make the following approximations using the limits as $N_S\to 0$ and $K\to\infty$:
\begin{align}
N_S^\prime &\approx (1-\eta)N_B,\\
(1+N_T)^{-K}&\approx e^{-N_S(1-\eta)N_B},\,{\text{and}}\\
\frac{A}{1-\mu}&\approx \frac{nM\eta N_S}{2(1-\eta)N_B}.
\end{align}
These lead to the following approximations for $p_c$ and $p_b$:
\begin{align}
\label{eq:pc_approx}p_c&\approx 1-\exp\left[-N_S\left(\frac{nM\eta\gamma}{2(1-\eta)N_B}+(1-\eta)N_B\right)\right]\\
\label{eq:pb_approx}p_b&\approx 1-\exp\left[-N_S(1-\eta)N_B\right],
\end{align}
where $\gamma=1-e^{-2\left(1+(1-\eta)N_B\right)}$.
Substitution of approximations in \eqref{eq:pc_approx} and \eqref{eq:pb_approx} into \eqref{eq:pe} and \eqref{eq:pd} yields:
\begin{align}
p_e&\approx \exp\left[-N_S(n-1)(1-\eta)N_B\right]\nonumber\\
\label{eq:pe_approx}&\phantom{\approx}-\exp\left[-N_Sn\left(\frac{M\eta\gamma}{2(1-\eta)N_B}+(1-\eta)N_B\right)\right]\\
&\approx\exp\left[-N_Sn(1-\eta)N_B\right]\nonumber\\
\label{eq:pe_approx_lb}&\phantom{\approx}-\exp\left[-N_Sn\left(\frac{M\eta\gamma}{2(1-\eta)N_B}+(1-\eta)N_B\right)\right],\\
p_d&\approx \exp\left[-N_Sn\left(\frac{M\eta\gamma}{2(1-\eta)N_B}+(1-\eta)N_B\right)\right]\nonumber\\
\label{eq:pd_approx}&\phantom{\approx}-\exp\left[-N_S\left(\frac{nM\eta\gamma}{2(1-\eta)N_B}+(1-\eta)N_B(n+1)\right)\right],
\end{align}
where we assume $n\gg1$ so that $n-1\approx n$ for the approximation in \eqref{eq:pe_approx_lb}.
When $N_S\to0$, we can approximate $p_e$ and $p_d$ by the Taylor series expansions at $N_S=0$ of \eqref{eq:pe_approx_lb} and \eqref{eq:pd_approx}, respectively:
\begin{align}
\label{eq:pe_approx_taylor}p_e&\approx \frac{N_SnM\eta\gamma}{2(1-\eta)N_B},\\
\label{eq:pd_approx_taylor}p_d&\approx N_S(1-\eta)N_B.
\end{align}
Substituting \eqref{eq:pe_approx_taylor} and \eqref{eq:pd_approx_taylor} into the last two terms of \eqref{eq:In}, and approximating $\frac{n-1}{n}\approx 1$, reveals that only the first term of \eqref{eq:In} has a significant dependence on $n$ in our regime of interest.
Thus, for the optimal order, we need:
\begin{align}
\label{eq:nopt}n^*&=\argmax_n\frac{p_e}{n}\log n.
\end{align}
The linear approximation in \eqref{eq:pd_approx_taylor} is insufficient to find $n^*$.  We follow the methodology in \cite{Jar17} by substituting in \eqref{eq:nopt} the quadratic Taylor series expansion at $N_S=0$,
\begin{align}
\label{eq:pe_approx_taylor2}p_e&\approx \frac{N_SnM\eta\gamma}{2(1-\eta)N_B}\nonumber\\
&\phantom{\approx}-\frac{N_S^2n^2M\eta\gamma\left(M\eta\gamma + 4(1-\eta)^2N_B^2\right)}{8(1-\eta)^2N_B^2}.
\end{align}
Let $u\equiv \frac{N_SM\eta\gamma}{2(1-\eta)N_B\ln 2}$ and $v\equiv\frac{N_S^2M\eta\gamma\left(M\eta\gamma + 4(1-\eta)^2N_B^2\right)}{8(1-\eta)^2N_B^2\ln 2}$. This reduces the problem in \eqref{eq:nopt} to finding the location of the extremal values of $f(n)=(u+vn)\ln n$ by solving
\begin{align}
\label{eq:dfdn}\frac{\dif f(n)}{\dif n}&=\frac{u}{vn}-1-\ln n=0
\end{align}
for $n$, which involves the principal branch of the Lambert $W$-function \cite[Sec.~4.13]{DLMF}:
\begin{align}
n^*&=\frac{u}{v}\left[W\left(\frac{u}{v}e\right)\right]^{-1},
\end{align}
where $W\left(xe^x\right)=x$ for $x\geq-1$.
Using equality $\ln W(x)=\ln(x)-W(x)$ for $x>0$ \cite[Eq.~4.13.3]{DLMF} and asymptotic expansion $W(x)=\ln(x)-\ln\ln (x)+o(1)$ as $\ln(x)\to\infty$ \cite[Eq.~4.13.10]{DLMF} in our regime of interest $N_S\to0$, we have:
\begin{align}
\log(n^*)&\approx\log\left(\frac{4(1-\eta)N_B}{N_S(M\eta\gamma+4(1-\eta)^2N_B^2)}\right)\nonumber\\
\label{eq:lognstar}&\phantom{\approx}-\log\left(\ln\left[\frac{4(1-\eta)N_B}{N_S(M\eta\gamma+4(1-\eta)^2N_B^2)}e\right]\right).
\end{align}
Substituting \eqref{eq:pe_approx_taylor} and \eqref{eq:lognstar} into \eqref{eq:REM}, we obtain:
\begin{widetext}
\begin{align}
\label{eq:REMapprox}R_E^{(M)}&\approx \frac{\eta N_S\gamma}{2(1-\eta)N_B}\left[\log\left[\frac{4(1-\eta)N_B}{N_S(M\eta\gamma+4(1-\eta)^2N_B^2)}\right]-\log\left[\ln\left[\frac{4(1-\eta)N_B}{N_S(M\eta\gamma+4(1-\eta)^2N_B^2)}e\right]\right]-g\left[\frac{2(1-\eta)^2N_B^2}{M\eta\gamma}\right]\right],
\end{align}
\end{widetext}
where $g(x)=(x+1)\log(x+1)-x\log x$. As $N_S\to0$, the logarithmic term dominates \eqref{eq:REMapprox}, and we obtain the scaling:
\begin{equation}
R_E^{(M)}=O\left(N_S\log \left(\frac{1}{N_S}\right)\right).
\label{eq:optimal_scaling}
\end{equation}

\subsection{Connection with PPM where dark-click rate is proportional to mean energy per slot}

In this subsection, we will consider a cruder approximation of $R_E^{(M)}$, providing an alternative proof of the scaling in~\eqref{eq:optimal_scaling}, but one that lets us establish a connection with a problem that was studied by Wang and Wornell in the context of coherent-state PPM modulation, where the dark click probability per mode $\lambda$ is proportional to the mean photon number per mode $\cal E$~\cite{Wan14}.

Recall that $R^{(M)}_E = {{\rm sup}_{n}R_E^{(M,n)}}$ is the envelope of capacities attained by our scheme over all $n$, for a given $M$. Applying the conditions pertinent to our problem setting, $\kappa N_S \ll N_S \ll 1 \ll N_B \ll 1/\kappa$, we get $N_S^\prime \to N_B$, $1/(1+N_T)^K \to e^{-N_SN_B}$ and $A/(1-\mu) \to nM\eta N_S/2N_B$, which lead to the following simplified asymptotic expressions: $1 - p_c \approx e^{-(n{\cal E}+\lambda)}$, and $1-p_b \approx e^{-\lambda}$, $\lambda = c{\cal E}$, with ${\cal E} = M\eta N_S/(2N_B)$ and $c = 2{N_B}^2/(M\eta)$ a constant. This is exactly the setting of $n$-mode coherent-state PPM modulation and direct detection, where the dark click probability per mode $\lambda$ is proportional to the mean photon number per mode ${\cal E}$~\cite{Wan14}. The leading-order terms of the optimal capacity for this setting, in the regime of ${\cal E} \ll 1$, is given by:
\begin{equation}
C_{\rm PPM}({\cal E}) \approx {\cal E}\log\frac{1}{\cal E} - {\cal E}\log \log\frac{1}{\cal E} - {\cal E}\log(1+c),
\end{equation}
with the optimal PPM order, $n = \lfloor \left({{\cal E}\log(1/{\cal E})}\right)^{-1} \rfloor$~\cite{Wan14}. Applying this result to our problem, we get
\begin{eqnarray}
{R^{(M)}_E} &\approx& \frac{1}{M}\left[\frac{M\eta N_S}{2N_B}\log\left(\frac{2N_B}{M\eta N_S}\right) \right. \nonumber \\
&&- \left. \frac{M\eta N_S}{2N_B}\log\left(\ln\left(\frac{2N_B}{M\eta N_S}\right)\right)\right].
\label{eq:REM_secondorder}
\end{eqnarray}
${R^{(M)}_E} \approx (\eta N_S/(2N_B))\log(2N_B/(M\eta N_S))$ to leading order. In the same regime as above, $\kappa N_S \ll N_S \ll 1 \ll N_B$, the leading order term for the Holevo capacity (attained using coherent states and Gaussian amplitude-and-phase modulation), $C \approx \eta N_S/N_B$, and that of the entangled-assisted capacity (achieved via an SPDC transmitter and phase-only modulation), $C_E \approx (\eta N_S/N_B)\log(1/N_S)$~\cite{Shi19}. It therefore follows that,
\begin{equation}
\frac{R^{(M)}_E}{C} \sim \log\left(\frac{1}{N_S}\right), \, \forall M,
\end{equation}
proving that our transmitter-receiver structure attains the optimal capacity scaling.

\subsection{Numerical comparisons}

\begin{figure}
\centering
\includegraphics[width=\columnwidth]{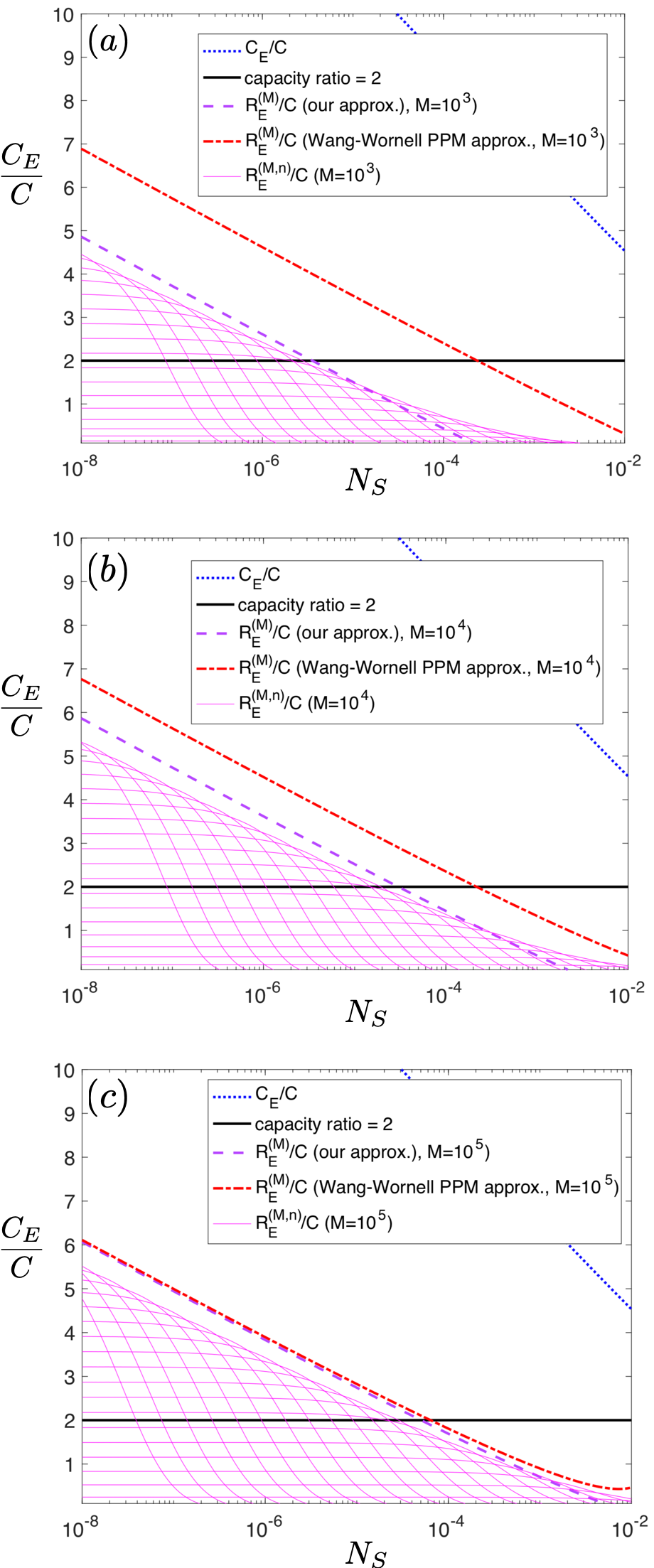}
\caption{Here, we plot ${R^{(M,n)}_E}/C$ for (a) $M=10^3$, (b) $M=10^4$ and (c) $M=10^5$, for $n \in \left\{2, 2^2, \ldots, 2^{20}\right\}$. We assumed $\eta = 0.01$ and $N_B = 10$ photons per mode, for all the plots. We compare the two approximations for the capacity-ratio envelope ${R^{(M)}_E}/C$: the one we obtained in Eq.~\eqref{eq:REMapprox} leveraging the Jarzyna-Banaszek analysis, and the one we obtained leveraging the Wang-Wornell analysis in Eq.~\eqref{eq:REM_secondorder}. It is seen that our approximation is tighter, especially for smaller $M$.}
\label{fig:capratio_approx_compare}
\end{figure}
In Fig.~\ref{fig:capratio_approx_compare}, we compare the two approximations for ${R^{(M)}_E}$: the one we obtained by modifying the Jarzyna-Banaszek analysis of PPM applied to our problem, shown in Eq.~\eqref{eq:REMapprox}, and the one we obtained from the Wang-Wornell PPM analysis, shown in Eq.~\eqref{eq:REM_secondorder}. It is seen that the former, our approximation, is closer to the true envelope, especially for smaller values of $M$.

\subsection{Optimum number of temporal modes in the phase-modulated SPDC pulse}

\begin{figure}
\centering
\includegraphics[width=\columnwidth]{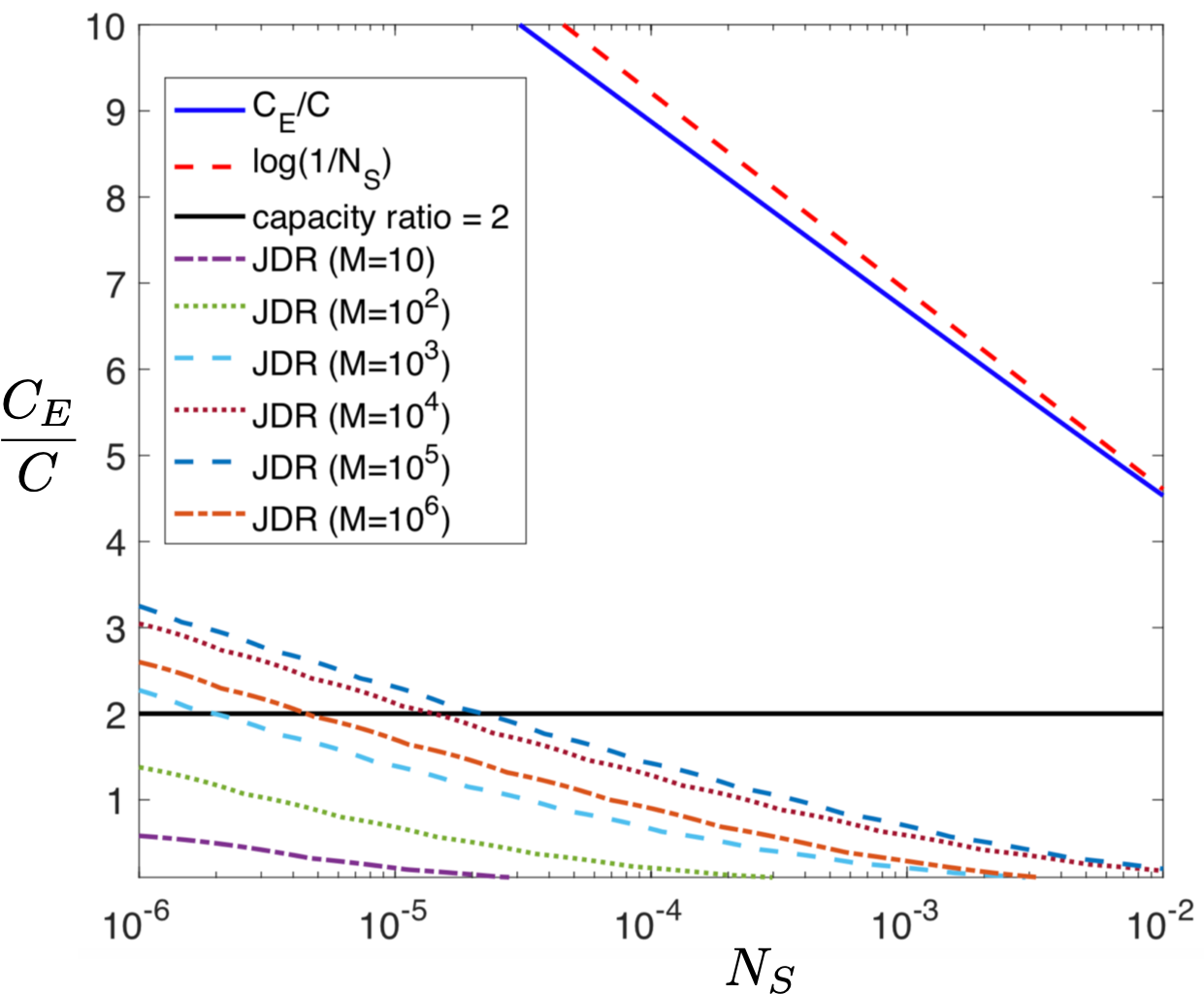}
\caption{Here we plot the envelopes of $R_E^{(M,n)}/C$ (taken over $n = 2, 4, 8, \ldots, 2^{14}$), for $M = 10, 100, \ldots, 10^6$. This shows that an optimum performance occurs at $M \sim 10^5$. We have assumed $\eta = 0.01$ and $N_B = 10$ photons per mode, for all the plots in this Figure.}
\label{fig:cap_ratios}
\end{figure}
In Fig.~\ref{fig:cap_ratios} we plot (the exact) $R^{(M)}_E$ as a function of $N_S$ for $M = 10, 10^2, \ldots, 10^6$. For the assumed values of $\eta = 0.01$ and $N_B = 10$ photons per mode used for the plots in this figure, the highest capacity occurs at around $M \sim 10^5$. The existence of such an optimum value of $M$ can be explained by the negative sign of the $M$-dependent second-order term in~\eqref{eq:REMapprox}.

For a typical SPDC entanglement source of optical bandwidth $W \sim 1$ THz, and $M \approx WT$, $M = 10^5$ modes in a signal pulse translates to a pulse duration of $T \sim 100$ ns. This means the BPSK phase-modulation bandwidth necessary would be $\sim 10$ MHz, which is readily realizable with commercial-grade electro-optical modulators at $1550$ nm.

\subsection{Potential improvements in joint detection receivers for future work}

In the main paper, we discuss a few ideas for improved receiver performance for entanglement assisted communications, including better codes (e.g., Reed Muller codes, along the lines of~\cite{Guh11a}) and exploiting detection of the ``noise modes" in the SFG stages, along the lines of~\cite{Zhu17}.

In addition, we would like to note that as $N_B->\infty$, $C_E/(C \ln N_S) \to 1$, while $R_E^{(M)}/(C \ln N_S) \to 1/2$, which indicates a possible check, to see if the entanglement-assisted capacity attained by an improved receiver design improves the aforesaid ratio from $1/2$ to $1$.

\section{OPA receiver analysis}
In the low photon number regime ($N_S\ll 1$) the communication capacities are well-approximated by the Taylor series expansion around $N_S = 0$.
For example, the Holevo capacity $C(\eta, N_S, N_B)$ is:
\begin{equation}
C(\eta, N_S, N_B) = \eta N_S \log\left(1+\frac{1}{(1-\eta)N_B}\right)+o(N_S).
\end{equation}
Here we derive the Taylor series expansion of the entanglement-assisted communication capacity with an SPDC source, BPSK modulation, and the OPA receiver~\cite{Guh09} of gain $G$.
We use it to evaluate the entanglement-assisted capacity gain achieved by an OPA receiver over the Holevo capacity. This channel's capacity is the classical mutual information between the random binary phase input $\theta\in\{0,\pi\}$, $P(\theta=0)=q$, modulating the block of $M$ transmitted symbols (i.e., $M$-fold tensor product of TMSV states) and the photon-count output $N$ of Bob's detector, optimized over the probability distribution of the input defined by $q$:
  \begin{align}
  C_{\text{EA-OPA}}(\eta,N_S,N_B)=\max_q I(\theta;N).
  \end{align}
The probability that the photon counter records $k$ photons over $M$ modes is: 
\begin{align}
\label{eq:Pk}P(k|\theta;M) = \frac{1}{(1+N_\theta)^M}{k+M-1 \choose k}\left(\frac{N_\theta}{1+N_\theta}\right)^k.
\end{align}
When phase $\theta$ is transmitted, the mean received photon number per mode is:
\begin{align}
N_{\theta} = G\ns+(G-1)N_S'+2C_p\sqrt{G(G-1)}\cos(\theta),
\end{align}
where $\ns$ is the mean photon number in each signal and idler mode, $\nt$ is the mean thermal noise injected by the environment, $\eta$ is the channel transmissivity, $N_S'\equiv \eta \ns+(1-\eta)\nt+1$, $G$ is the gain of the OPA, and $C_p\equiv \sqrt{\eta \ns(\ns+1)}$.
	
The Taylor series of mutual information $I(\theta;N)$ at $\ns=0$ is:
\begin{align}
\nonumber I(\theta;N)=-\ns\sum_{k=0}^{\infty}\sum_{\theta\in\{0,\pi\}}\left.Q_\theta(k,\ns)\right|_{\ns=0}+o(N_S),
\end{align}
where 
\begin{widetext}
\begin{align}
Q_\theta(k,\ns) =  
	\begin{cases} 
      q\frac{\d\atermM}{\d{\ns}}\log\left(q+(1-q)\frac{\btermM}{\atermM}\right), & \theta = 0 \\
      (1-q)\frac{\d\btermM}{\d{\ns}}\log\left((1-q)+q\frac{\atermM}{\btermM}\right), & \theta = \pi 
   \end{cases}.
\end{align}
\end{widetext}
Substitution of \eqref{eq:Pk} and evaluation of 
  $Q_\theta(k, \ns)\big\rvert_{\ns = 0}$ by taking the limit 
  $\lim_{\ns \to 0} Q_\theta(k,\ns)$ yields:
\begin{widetext}
\begin{align}
I(\theta;N)=&\ns\sum_{k=0}^{\infty}8q(1-q){\eta}G(G-1)^{k-1}(N_B')^{k-2}(G+(1-\eta)(G-1)\nt)^{k-M-2}(k+(G-1)M(N_B')^2{k+M-1 \choose k}\nonumber \\ & 	+ o(\ns),\label{eq:sumI}	
\end{align}
\end{widetext}
where $N_B'\equiv1+(1-\eta)N_B$.
Well-known results for the moments of binomial distribution are used to evaluate the sum in \eqref{eq:sumI}. Maximizing over $q$ yields:
\begin{align}
C_{\text{EA-OPA}}(\eta,N_S,N_B)=&\frac{2{\eta}GM\ns}{N_B'(G+(1-\eta)(G-1)\nt)}\nonumber \\&+ o(\ns).
\end{align}

The maximum gain from using the SPDC source, BPSK modulation and the OPA receiver over the Holevo capacity when $N_S\ll 1$ and $N_B\gg 0$ is thus:
\begin{align}
\lim_{G \downarrow 1}\lim_{\nt \to \infty} \frac{C_{\text{EA-OPA}}(\eta,N_S,N_B)}{M\times C(\eta,N_S,N_B)} = 2,
\end{align}
where $\lim{G\downarrow1}$ indicates a one-sided limit taken from above, and we normalize the denominator by $M$ to account for employing block encoding of $M$ symbols.
We note that, with such normalization, the gain does not depend on $M$.  There is also no dependence on the transmissivity~$\eta$.

\section{PPM and OOK modulation for Entanglement-Assisted Communications}


In this Appendix, we will discuss alternative modulation formats for entanglement-assisted communications, which also leverage continuous-variable SPDC-based pre-shared entanglement, and can also achieve the $\log(1/N_S)$ capacity-ratio improvement over the Holevo capacity.

\subsection{Pulse position modulation (PPM)}
At the $n$ output modes of the $K$ Green Machine (GM) circuits in Fig.~1 of the main paper, the state of the $nK$ output modes resembles pulse-position modulation (PPM): One block of $K$ modes carries displaced thermal states ${\hat \rho}_{\rm th}(\sqrt{n N_k}, N_T)$, where $N_k = M\kappa \eta N_S(1+N_S)\mu^{k-1}$, $1 \le k \le K$, with $\mu = (1-\kappa(1+N_S^\prime))^2$, $N_T = \kappa N_S N_S^\prime$, $N_S^\prime = \eta N_S + (1-\eta)N_B$. The remainder $n-1$ of the $K$-mode blocks are excited in zero-mean thermal states ${\hat \rho}_{\rm th}(0, N_T)$.

One alternative to the aforesaid scheme described in the main paper is for Alice to directly modulate PPM codewords. In such a scheme, Alice and Bob would need to pre-share (brighter) SPDC signal-idler mode pairs of mean photon number per mode $nN_S$, and Alice would send an $M$-temporal mode signal pulse (of mean photon number $nN_S$) and nothing (vacuum) in $n-1$ pulse slots. So, only $M$ modes will be occupied by signal pulses out of each $Mn$ transmitted modes. FF-SFG stages will be used to demodulate, as before, but no GM stages will be used. The state of the $nK$ output modes of the $n$ $K$-stage FF-SFG modules will be identical to the above: One block of $K$ modes carries displaced thermal states ${\hat \rho}_{\rm th}(\sqrt{n N_k}, N_T)$, and the remainder $n-1$ of the $K$-mode blocks will be excited in zero-mean thermal states ${\hat \rho}_{\rm th}(0, N_T)$. 

The mean transmit photon number of both schemes are identical. The DMC induced by the modulation-code-receiver combination for both schemes are identical. Hence, the capacity achieved by the two schemes are identical. The optimal PPM order for the second scheme is the optimal Hadamard-code length for the first scheme. That optimal PPM-order (or Hadamard code length) is given by: $n \sim ({\cal E}\log(1/{\cal E}))^{-1}$ with ${\cal E} = M\eta N_S/(2N_B)$, which translates to $nN_S \sim \frac{N_0}{\log(N_0/N_S)}$ with $N_0 = 2N_B/(M\eta)$. For the numbers in Fig.~\ref{fig:cap_ratios}, i.e., $\eta=0.01$, $N_B=10$, $M=10^4$, we get $N_0 = 0.2$, and optimal $n \approx 7$. This implies that that for $N_S < 0.01$, $nN_S \lesssim 0.07$. This means that the idler pulses are still in the regime that the implicit ``qubit approximation" analysis of the SFG borrowed from~\cite{Zhu17} is valid. 

There are key operational differences however, between the two schemes, which are described below:
\begin{enumerate}
\item {\textbf{Peak power usage}}---Even though the mean photon number that is transmitted over the channel is identical for both schemes, the peak power is not. The PPM scheme uses $n$ times more peak power than the BPSK scheme. For the above said numbers, the optimal PPM order $n \approx 7$, which implies the peak power is $7$ times that of BPSK. However, the BPSK scheme is slightly more restrictive since Hadamard codes exist only for $n$ that is an integer power of $2$. But, it is possible to redesign the BPSK scheme with complex-valued Hadamard codes that would work for all integer $n$.
\item {\textbf{Entanglement consumption}}---More important than the peak power advantage the BPSK scheme enjoys is that its entanglement consumption is lower. Despite the fact that the mean photon number per transmitted mode is $N_S$ for both schemes, in the PPM scheme, every $M$-mode SPDC pulse that needs to be pre-shared must have $nN_S$ photons per mode. This is true, even though $(n-1)/n$ fraction of the signal pulses of the pre-shared entangled states will never be transmitted in the PPM scheme. This is a major drawback for this scheme.
\item {\textbf{Receiver complexity}}---The BPSK scheme needs the $K$ Green Machine circuits, in addition to the FF-SFG modules. That is an added receiver complexity for the BPSK scheme over the PPM scheme.
\item {\textbf{Using the noise modes of FF-SFG stages}}---In the BPSK scheme described in the main paper, we ignored the $nK$ ``noise modes", labeled ${\hat e}_k^{(i)}$, shown in Fig.~1 of the main paper. In the operational regime relevant to our problem, for both the BPSK and PPM schemes, the state of mode ${\hat e}_k^{(i)}$ is close to zero-mean thermal states, of the same mean photon number as that of the corresponding sum-frequency mode, ${\hat b}_k^{(i)}$. The capacity analyses (for both BPSK and PPM) above ignores these modes. There is information about the transmitted codeword in these noise modes, which can only increase the achieved capacity. For the PPM scheme, one can simply do photon counting on all the ${\hat e}_k^{(i)}$ modes. For PPM, for the pulse-containing block of $K$ noise modes ${\hat e}_k^{(i)}$, $1 \le k \le K$, simple on-off direct detection of those modes effectively make the ``on" pulse of the PPM twice the energy, causing the capacity-ratio plots to be shifted to the right by $\log_10(2)$. This is a small improvement, but one that only requires additional single-photon detectors to obtain. To obtain a similar capacity improvement for the BPSK scheme leveraging the ${\hat e}_k^{(i)}$ modes, one will need to employ a feedback-based scheme similar to the one in~\cite{Zhu17}, where based on photon-detection events at the noise modes, one will need to apply adaptively two-mode squeezing before and after each of the SFG stages within the FF-SFG modules. A rigorous analysis of this will require a second-order analysis of SFG, because the photo-detection statistics across the ${\hat e}_k^{(i)}$ modes are correlated.
\end{enumerate}

\subsection{On-off keying (OOK)}
Finally, PPM can be thought of as a modulation code over an on-off keying (OOK) alphabet, and hence its capacity is strictly inferior to that of OOK, although it is very close to OOK when ${\cal E} \ll 1$. This means that an OOK version of our modulation format will also work to attain the $\log(1/N_S)$ capacity ratio. Here, the ``on" symbol (transmission of the $M$-mode signal pulse) will be associated with a prior probability $p$ and the ``off" symbol (no signal transmission) with a prior probability $1-p$, with $p \sim {\cal E}\log(1/{\cal E})$ assuming the role of the inverse-order $1/n$ of PPM, except that there is now no restriction that there must be exactly one ``on" pulse in every $n$-pulse block. 

Despite the fact that the same entanglement-assisted capacity could be attained with PPM and OOK modulation formats as can be with our BPSK scheme---that uses the GM stages in addition to the FF-SFG modules---the latter may be preferable in practice. This is because both PPM and OOK modulation formats will require Alice and Bob to pre-share more entanglement, i.e., the pre-shared signal-idler mode pairs will have to have a higher mean photon number per mode, and fault-tolerant entanglement distribution to pre-share the resource necessary for supporting entanglement-assisted communications, is likely to be the most expensive part of the process in a future practical implementation. 

\end{document}